\documentstyle[12pt,aasms4,epsf]{article}
\received{00  2001}
\accepted{00 2001}

\def\ee #1 {\times 10^{#1}}          
\def\ut #1 #2 { \, \textrm{#1}^{#2}} 
\def\u #1 { \, \textrm{#1}}          

\def\kms    {\hbox{km{\hskip0.1em}s$^{-1}$}}    
\def\msol   {\hbox{$M_\odot$}}                  

\def\kms    {\hbox{km{\hskip0.1em}s$^{-1}$}}    
\def\etal   {{\it et al. }}                     

\begin{document}
\title{The  Origin of X-ray Emission from a Galactic Center Molecular Cloud: Low
Energy Cosmic Ray Electrons}

\author{F. Yusef-Zadeh\footnote
{Department of Physics and Astronomy, Northwestern University,
Evanston, Il. 60208 (zadeh@northwestern.edu)}, C.
Law\footnote{Harvard-Smithsonian Center for Astrophysics, Cambridge,
MA. 02138
 (claw@head-cfa.harvard.edu)}
 \& M. Wardle\footnote{Research Centre for Theoretical Astrophysics, 
School of Physics,
University of Sydney, NSW 2006, Australia
(wardle@physics.usyd.edu.au)}}

\begin{abstract}

	The Galactic center region near l$\approx0.2^0$ hosts a mixture of
nonthermal linear filaments and thermal radio continuum features associated
with the radio Arc.  Chandra observations of this region reveal an X-ray
filament and diffuse emission with an extent of roughly 60$''\times2''$ and
5$'\times3'$, respectively. The X-ray filament lies at the edge of the
nonthermal radio filaments and a dense molecular shell G0.13-0.13 that has
an unusually high kinetic temperature $\ge$70K.  These observations
demonstrate that the G0.13--0.13 molecular cloud and the nonthermal radio
filaments of the Arc are interacting.  The diffuse X-ray emission is
correlated with the molecular shell and is fitted either by two-temperature
(1 and 10 keV) thermal emission or by power-law and 1 keV thermal gas.
Fluorescent 6.4 keV line emission is also detected throughout the molecular
shell.  This cloud coincides within the error circle of a steady
unidentified EGRET source 3EGJ1746--2851.

We argue that low-energy cosmic ray electrons (LECRe) produce the power-law
continuum by bremsstrahlung and 6.4 keV line emission from the filament
and the diffuse cloud with the implication on the origin of the Galactic
ridge X-ray emission.  The strong 6.4 keV Fe-line emission seen from other
Galactic center clouds could be produced in a similar fashion rather than
via fluorescent emission induced by a transient hard X-ray source in the
Galactic center.  In addition, heating by ionization induced by LECRe are
also responsible for the high temperature of G0.13-0.13.  The gamma-ray
source is a result of brehmsstrahlung by the high-energy tail of the
electron energy distribution.

\end{abstract}



\section{Introduction}

The Galactic center radio continuum Arc is a complex system of thermal
and nonthermal features distributed near l$\approx0.2^0$.  The
nonthermal components of the Arc -- the linear filaments -- are long
($>15'$), narrow ($\approx5''-10''$) linearly polarized synchrotron
emitting structures tracing organized magnetic fields which run
perpendicular to the Galactic plane (Yusef-Zadeh, Morris and Chance
1984; Tsuboi \etal 1985; Sofue \etal 1986; Yusef-Zadeh and Morris
1987; Inoue \etal 1989; Reich \etal 2000).  Recently, Tsuboi, Ukita
and Handa (hereafter TUH; 1997) and Oka \etal (2001) have detected a
CS and CO molecular cloud with a size of about 3$'\times4'$ centered
at G0.13-0.13 or G0.11-0.11, respectively.  The G0.13-0.13 molecular
cloud is about 7 pc in diameter (assuming a distance of 8 kpc) and is
distinguished from other clouds within the inner 30$'$ of the Galactic
center by exhibiting an enhanced kinetic temperature T$_k\ge$70K, and
relative isolation from the rest of clouds in the Galactic center (Oka
\etal 2001).  A morphological argument has been made for a site of
interaction between a dense molecular shell G0.13-0.13 and the
neighboring NTFs (TUH; Oka \etal 2001).

Here we present high resolution Chandra observations of the Galactic
center radio Arc and report the detection of enhanced diffuse continuum
and 6.4 keV line emission as well as filamentary X-ray emission from
G0.13-0.13.  Earlier ASCA X-ray observations suggested the presence of
diffuse, hot (10 keV) X-ray gas throughout this region (Koyama et al.
1996).  We present an alternative power-law and thermal fit to the
diffuse X-ray gas in G0.13-0.13.  In addition the power-law component
and strong Fe 6.4 keV line are consistent with bremsstrahlung and
collisional ionization by low-energy cosmic-ray electrons.  The
implication of such a mechanism to the origin of the hard X-ray
emission from the Galactic ridge is also discussed.  The associated
heating explains the high kinetic temperature of the G0.13-0.13
molecular cloud.  This is the strongest evidence yet of the
interaction with NTFs, leading us to identify the Galactic center
EGRET source 3EGJ1746-2851 (Hartman \etal 1999) with the X-ray feature
and G0.13-0.13 molecular cloud.

\section{Observations and Results}

An account of Chandra observations of the Galactic center Arc has been
given earlier (Yusef-Zadeh \etal 2001).  The X-ray spectrum was
extracted from an almost-rectangular region, as shown in Figure 1.  We
used Sherpa to fit the extracted spectrum with either: (1) an absorbed
power-law plus thermal model, using the MEKAL model for the thermal
component (Mewe \etal 1985), or (2) using an absorbed, two temperature
model.  Each component of the two temperature fit was permitted to
have different temperatures and volume emission measures, while the
elemental abundances (Si, S, Ar, Ca, and Fe) were constrained to be
identical for both components.  For both both models (1) and (2), we
added a Gaussian component fixed at 6.4 keV, to account for
fluorescent Iron K$\alpha$ line emission.  Diffuse X-ray emission
pervades the galactic center region making the extraction of a
non-xray background nearly impossible.  We accounted for the non-xray
background by using the CXC's ``blank-sky'' fields.  The background
spectrum was extracted from the same region as the source spectrum, in
order to avoid any possible response-dependent affects.

An adaptively smoothed X-ray image of a region near G0.13-0.13 (right
panel of Figure 1) shows diffuse X-ray emission distributed
prominently in a rectangular area of $\sim5.3'^2$.  The
full-resolution X-ray image of the same region (left
panel) reveals a linear X-ray filament with a size of
$60''\times2''$.  A compact source is also detected toward the
filament at (epoch 2000) $\alpha=17^h 46^m 21.56^s, \delta=-28^0 52'
56.8''$.  The enhanced diffuse and filamentary X-ray gas coincides
with the CS and CO cloud G0.13--0.13 (TUH and Oka \etal 2001).  The
box drawn on Figure 1 shows the extent of G0.13-0.13 CS cloud.  Figure
2 shows contours of CS (1--0) emission from the molecular cloud shell
G0.13-0.13 with velocities in the range V$_{LSR}=$15 to 45 \kms (TUH).
The contours are superimposed on a low-resolution X-ray distribution
between 0.5 and 9 keV, which is displayed in greyscale.  There is a
strong spatial correlation between the diffuse X-ray gas and the CS
line emission.  The CS cloud is limb brightened and the X-ray images
of Figure 1 show a strong X-ray filament to the north tracing the edge
of the CS cloud.  The estimated mass of the high density CS and the CO
clouds are 2--3.6$\times10^5$ and $10^6$ \msol, respectively.

TUH argue that the northwestern limb of the CS molecular shell with a
density of 6$\times10^4$ cm$^{-4}$ is dynamically interacting with the
Galactic center radio filaments because of the distorted structure as
well as the large linewidth of the CS cloud, which exceeds 40 \kms.
The enhanced kinetic temperature of the CO cloud G0.13-0.13 plus the
abrupt velocity of the CS cloud at 40 \kms provide additional
support to the idea that this is an unusual cloud and is possibly
interacting with the NTFs (Oka \etal 2001).  The X-ray and molecular
gas distributions noted in Figure 2 also coincide with a steady source
of strong $\gamma$-ray emission between 30 MeV and 10 GeV as detected
by EGRET on board the {\it{Compton Gamma Ray Observatory}}.  The cross
in Figure 2 shows the position of 3EGJ1746--2851 (Hartman \etal 1999)
with the error bar containing the 95\% confidence contour.  The photon
index is $\alpha$=1.7$\pm0.7$, and its flux is estimated to be
1.2$\times10^{-6}$ photons cm$^{-2}$ s$^{-1}$ with energies greater
than 100 MeV, corresponding to $10^{40}$ photons s$^{-1}$ at the
distance of the Galactic center.

Figure 3 shows two different images of the fluorescent emission (Fe
K$\alpha$) at 6.4 keV. The left panel shows the fit value for the
Gaussian normalization to each of these spectra and the right panel
shows an adaptively smoothed 6.4 keV image.  In general, the images
correlate well, lending support to each method and in turn the
correlation of 6.4 keV emission with the CS(1--0) contours.  In
particular, we note that each method shows three extended ``blobs'' in
the SW edge of the CS cloud with strong 6.4 keV line emission.  These
blobs with a size of $\approx1'$ do not show any obvious relation to
point sources in the full-resolution data.  The low-resolution 6.4 keV
image produced from ASCA observations show 6.4 keV emission from this
region (Koyama \etal 1996).

Figure 4  shows contours of the X-ray filament superimposed on the
$\lambda$20cm continuum image of the NTFs. The X-ray filament appears to
be situated at the outer boundary but parallel to a number of small radio
filaments having the same dimension as the X-ray filament.
The nearest radio filament is  displaced by $\sim15''$ from
the X-ray filament but there is diffuse $\lambda$90cm emission at a level
of 200 to 245 mJy per 41.7$''\times22.7''$ beam area coincident with the
X-ray filament (Anantharamaiah \etal 1991).

A spectrum shown in Figure 5 is extracted from the rectangular region of
diffuse X-ray
gas which is spread over four chips of the detector.  This  spectrum
cover the densest region of the CS cloud G0.13-0.13.  The point
sources have been removed before the spectrum is fitted.  The spectrum
of the diffuse X-ray gas is equally-well fitted by two different
models.  One is thermal bremsstrahlung with two temperatures at 0.97
and 10.5 keV and a 6.4 keV Gaussian corresponding to the fluorescent
Fe K/$\alpha$ emission with column density N$_H\sim5.6\times10^{22}$
cm$^{-2}$.  The thermal flux of absorption-corrected spectrum
integrated over the box shown in Figure 1, between 0.5 and 10 keV give
2.7$\times10^{-11}$ and 1.8$\times10^{-11}$ ergs cm$^{-2}$ s$^{-1}$
for the low and high temperature components, respectively.
Alternative spectral fitting using weighted response files was done
with a combination of a power-law with a photon index $\alpha=1.38$, a
single temperature of 0.97 keV and a 6.4 keV Gaussian line with
N$_H\sim6.8\times10^{22}$ cm$^{-2}$.  The flux of absorption-corrected
spectrum for 0.97 keV component and the power law component give
2.3$\times10^{-10}$ and 1.6$\times10^{-11}$ ergs cm$^{-2}$ s$^{-1}$,
respectively.  The 6.4 keV line emission in both models give
1.2$\times10^{-12}$ ergs cm$^{-2}$ s$^{-1}$.  The fitted values are
shown in Table 1.  Both models give equally well-fit spectra with the
value of $\chi^2\approx1.2$.  We also note a strong corrleation
between the amplitude of the power-law fit and the strength of 6.4 keV
line emission which in turn is correlated spatially with the CS line
emission.  The spectrum of the X-ray filament which is included in the
model fits could not easily be isolated from the contribution from the
background diffuse gas.  Using different areas around the filament as
well as including the filament and the surrounding diffuse cloud, the
extracted spectra were fit equally well either by a power law or
thermal model and showed similar characteristics to that of the
surrounding diffuse gas cloud with $\alpha$ ranging between 1.3 and
1.4.  The only difference in the two fitted spectra is the apparent
overabundance of Fe with large error.  The contribution of 6.4 keV
line which is not resolved from 6.7 keV emission may be responsible
for the large error.

\section*{Discussion}

The spectrum of the diffuse X-ray emission coincident with the
G0.13--0.13 molecular cloud can be satisfactorily modeled as emission
from a two-component thermal plasma with temperatures of $\sim 1$ and
10 keV. Although similar two-temperature models have been invoked
previously within the Galactic center region (Koyama \etal 1996), the
production and confinement of the inferred hot component within the
Galactic center region is problematic.

This is also true of the Galactic X-ray ridge emission, but recently
it has been demonstrated that bremsstrahlung from low-energy
cosmic-ray electrons more naturally explains the spectrum below 10 keV
and the strength of the Fe 6.4 keV line (Valinia \etal 2000).  The
nonthermal spectrum below 10 keV is consistent with the power law
component detected above 10 keV (Yamasaki \etal 1997; Valinia \&
Marshall 1998).  Ginga observations show a hard tail in the spectrum
of the inner region  of the Galactic center with a power law like spectrum.
(Yamauchi et al.  1990; Yamasaki et al.  1996, 1997).

Here we show that this model also applies to G0.13--0.13.  First, we
note that the absorption-corrected spectrum is well-fit by a soft
thermal component plus a power-law.  The photon index of $\approx
1.38$ is consistent with that expected from bremsstrahlung by low
energy ($\la 100$ keV) cosmic-ray electrons, $-1.3$ to $-1.4$ over the
range 1--10 keV (Valinia et al 2001).  The calculations of Valinia et
al yield a photon production rate per H nucleus of $1.4\times10^{-20}$
U( eV cm$^{-3}$ ) ph s$^{-1}$  H$^{-1}$ at 1 keV, and a
production of 6.4 keV Fe line photons of $4.8\times10^{-22}$ U( eV
cm$^{-3}$ ) ph s$^{-1}$  H$^{-1}$, where U is the energy
density of low energy cosmic-ray electron.  Adopting a total mass of
G0.13--0.13 of $10^6$ \msol and a distance of 8 kpc, our measured
fluxes of the power-law component at 1 keV and in the 6.4 keV line
correspond to cosmic-ray energy densities of 1.1 and 1.9 $\u eV \ut cm
-3 $ respectively.  Solar abundances are assumed in Valinia et al's
calculation, so an iron abundance of twice solar would reconcile the
continuum and iron-line fluxes.  Although our fit to the thermal
component (Table 1) gives a somewhat higher iron abundance, it is
poorly determined and in any case could differ from the iron abundance
in the cold cloud.  Within the uncertainties in the shape of the
electron spectrum, the model reproduces the observed ratio of 6.4 keV
line to continuum, and requires a reasonable energy density in
cosmic-ray electrons.  In comparison, we note that Valinia et al
(2000) estimate that $U\sim 0.2 \u eV \ut cm -3 $ would be sufficient
to explain the galactic ridge emission.

The hypothesized low energy electron population would be the dominant
source of ionization and heating within the G0.13--0.13 molecular
cloud, explaining in particular the high kinetic temperature ($\sim 70
\u K $) of G0.13--0.13 (Oka et al 2001) and possibly the high abundance of
SiO emission in the vicinity of the Arc correlating with the
distribution of the 6.4 keV line emission (Martin-Pintado \etal 1997).
The ionization cross section for H$_2$ by electrons is $2\ee -17 \ut
cm -2 [E(\u keV )]^{-1}$ (e.g.  Voit 1991), and for the electron
spectrum of Valinia et al (2000), the ionization rate is $\sim
2\times 10^{-14}\ut s -1 \ut H -1 $, 1000 times the value for molecular clouds
in the solar neighborhood.  The effect on the cloud can be estimated
using the calculations of Maloney, Hollenbach \& Tielens (1996; see
their eq [15] and Fig 3a) for ionization and heating by X-rays, as the
effects are similar.  Their results indicate a fractional ionization
of $\sim 10^{-6 }$ and a gas temperature of $\sim 50$--$100 \u K $.
The enhanced abundance of SiO gas can be explained perhaps by the
impact of low energy cosmic rays with dust grains.

The X-ray filament has a similar spectrum to the diffuse emission
associated with G0.13--0.13, and so is likely to be produced by the
same mechanism.  In this scenario, the emission (both in continuum and
the 6.4 keV line), which is enhanced by $\sim 1.5$ times that from
G0.13--0.13, is generated in a volume of the cloud containing one
percent of it's mass, $\sim 10^4 \msol$ of material.  The required
energy density is therefore increased to $\sim 150\u eV \ut cm -3 $,
still small compared to the $\sim 3\ee 4 \u eV \ut cm -3 $ of a 1 mG
magnetic field.  A plausible source of the enhanced cosmic-ray
electrons is acceleration at the interaction site between the
G0.13--0.13 cloud and the nonthermal filaments radio filaments of the
Arc(TUH; Oka \etal 2001).  The distribution of diffuse and filamentary
X-ray gas lying at the edge of the NTFs imply the presence of
nonthermal high energy electrons in the vicinity of the diffuse X-ray
source.  The X-ray filament being distributed parallel to the NTFs is
strongly suggestive of an interaction of X-ray gas with nonthermal
filaments.  In particular the displacement of the X-ray and radio
filaments from each other suggest that the high-energy component of
the cosmic ray electrons produce synchrotron radio emission from
highly organized magnetic field whereas the low-energy cosmic ray
electrons produce bremsstrahlung emission from the edge of the CS
G0.13--0.13 cloud.  The local cloud material is heated to $\sim
3000$K, and is predicted to be predominantly atomic with a fractional
ionization $\sim 10^{-2}$ (Maloney et al 1996).

The G0.13--0.13 cloud and the diffuse and filamentary X-ray features lie
within the 95\% error circle of the unidentified EGRET source 3EGJ1746--2851.
This suggests that the source arises as a result of the interaction of
the filament with the cloud.  The $E^{-1.7}$ photon spectrum of
3EGJ1746--2851 matches onto the cloud spectrum at about 10 keV when
extended down to X-ray energies.  This suggests that the EGRET source
is also produced by brehmsstrahlung, and that the electron spectrum
extends up to GeV energies (to $E_{\mathrm{max}}$, say), with an
$E^{-1.7}$ dependence in this range.  The required energy density in
high-energy electrons is $\approx 30 E_{\mathrm{max}}/(100\u GeV ) \u
eV \ut cm -3 $.  This model is supported by the spectral index
measurements of the nonthermal filaments between cm and mm
wavelengths.  The spectral indices ($p$, where $S_\nu \propto \nu^p$)
are positive along the vertical filaments of the Arc but there is an
anomalous filament near G0.16-0.15 with $p\sim-0.35$ (Anantharamaiah
\etal 1992), consistent with synchrotron emission from GeV electrons
with an $E^{-1.7}$ spectrum.  Close inspection of the spectral index
map of Anantharamaiah \etal indicates a value similar to that of the
anomalous filament near the X-ray filament.  Future high resolution
spectral index measurements of this region should give a more accurate
value of $p$ in this unusual region.

We have shown that the strong Fe 6.4 keV line emission from the
G0.13--0.13 cloud can be explained by impact ionization by low-energy
cosmic-ray electrons.  A similar interpretation may well be applicable
to the diffuse line emission detected from other galactic center
molecular clouds, notably Sgr B2 and Sgr C and througout the inner two
degrees of the Galactic center (e.g.  Murakami \etal 2001a,b; Wang,
Gothelf and Lang 2002).  This has been interpreted in terms of
fluorescent emission resulting from irradiation by an intense source
of hard X-rays (Sunyaev, Markovitch and Pavlinsky 1993; Koyama \etal
1996).  As no such source is apparent, it has been hypothesised that a
transient source, perhaps associated with Sgr A$^*$ is responsible.
However, a low-energy cosmic-ray electron population permeating the
galactic center region would explain this.  The geometry of the 6.4
keV emission from the NE edge of the CS cloud is not consistent with a
powerful flare from the Galactic center.  The excess of supernova
remnants detected in the Galactic center region (Gray 1994) is likely
to be responsible for enhancing the flux of high energy cosmic ray
particles in the central regions of the Galaxy.



\begin{figure}
\caption{
(left panel)
Full-resolution image of the filtered x-ray data.
The
region was chosen to fully include the CS (1--0) emission from
G0.13-0.13 (Tsuboi \etal 1997).  The filament is visible near the
northeast border of the region.  (right panel)  Adaptively-smoothed,
exposure-corrected image of the X-ray emission with identical border
as in the left  panel. }

\end{figure}

\begin{figure}
\caption{Contours of CS (1--0) molecular gas (Tsuboi \etal 1997)
integrated over the velocity range between 15 and 45 \kms superimposed on
the smoothed distribution of X-ray gas between 0.5 and 9 keV. The cross
coincides with
the error bar containing the 95\% confidence contour of
3EGJ1746-2851 (Hartman \etal 1999). }
\end{figure}

\begin{figure}
\caption{(left panel)
The grayscale shows the distribution of the
normalization of the 6.4 keV gaussian fit to each of the 81 (9$\times$9),
50$''$ spectra.  To fit the spectra, a template spectrum was extracted from
the
rectangular region shown in figure 1.  The normalization
of the
gaussian line and continuum  was allowed to
vary.
The grayscale range is between  1$\times10^{-8}$ and 9.5$\times10^{-6}$
photons cm$^{-2} s^{-1}$ per 50$^2$'' pixel.
The fits
are restricted to energies from $\approx$4--9 keV in order to avoid
the more variable, thermally dominated portions of the spectra from
1--4 keV. (right
panel) The grayscale shows an adaptively smoothed image of all X-ray
photons with energy from 6.15-6.65 keV, after removing the point
sources.  The contours show the distribution of the CS(1-0) emission.
The grayscale range is between  4$\times10^{-11}$ and 1.3$\times10^{-9}$
photons cm$^{-2} s^{-1}$ per 0.5$^2$'' pixel.}
\end{figure}

\begin{figure}
\caption{Contours of X-ray filament with a
spatial resolution of
2.5$''\times2.5''$ are superimposed on a $\lambda$20cm VLA image of the
nonthermal
filaments of the Arc with a resolution of 10$''$.}
\end{figure}

\begin{figure}
\caption{Background subtracted, ACIS spectrum of the CS-emitting
region as shown in a box in Figure 1.  Solid line shows the fit of the
absorbed
1-temperature/power-law plus 6.4 keV Gaussian source model.This figure
is shown only in electroinic version of this paper.}
\end{figure}

\end{document}